\documentclass[aps,amsmath,amssymb,prl,preprint]{revtex4}

\usepackage{graphicx}

\begin{document}

\author{N. Olver and I.V. Barashenkov}

\affiliation{Department of Mathematics, University of Cape Town,
 Rondebosch 7701, South Africa}

\title{Complex sine-Gordon-2:
a new algorithm for  multivortex solutions on the plane}

\begin{abstract}
We present a new vorticity-raising transformation for the second integrable
complexification of the sine-Gordon equation on the plane. The new
transformation is a product of four Schlesinger maps of the
Painlev\'{e}-V to
itself, and allows a more efficient construction of the $n$-vortex solution
than the previously reported transformation comprising a product of
$2n$ maps.

\end{abstract}

\vspace{10mm}


\vspace{1cm}

\date{\today}
\maketitle

\newpage

\textbf{1.} The complex sine-Gordon equation, also known as the
Lund-Regge model,
was introduced in 1970s in several field-theoretic contexts
\cite{Pohlmeyer,Lund_Regge,G1,Neveu_Papanicolaou}.
In $(2+0)$-dimensional space, the equation assumes the form
\begin{equation}
 \nabla^2 \psi +
 \frac{(\nabla \psi)^2 \, { \overline \psi}}{1-|\psi|^2}
 + \psi (1-|\psi|^2)=0.
 \label{csG-1}
\end{equation}
 (Here and below,
$\nabla={\bf i} \partial_x + {\bf j} \partial_y$.) We will be
referring to eq.\eqref{csG-1} as the complex sine-Gordon-1, in order to
distinguish it from another integrable complexification of the sine-Gordon
theory, the so-called complex sine-Gordon-2:
\begin{equation}
\nabla^2 \psi + \frac{(\nabla\psi)^2\overline{\psi}}{2-|\psi|^2} +
\tfrac{1}{2}\psi(1-|\psi|^2)(2-|\psi|^2)=0.
\label{eq:csg2eqmotion}
\end{equation}
This model has also been known since the late 1970s,
yet in  $(1+1)$-dimensional space \cite{Sciuto,G2}. The names stem
from the fact that if we assume that $\psi$ is real and substitute
$\psi=\sin(\alpha/2)$ in \eqref{csG-1} and $\psi = \sqrt{2}\sin(\alpha/4)$ in
\eqref{eq:csg2eqmotion}, both systems reduce to the conventional, real,
sine-Gordon equation $\nabla^2 \alpha + \sin \alpha = 0$. In
the physics literature, it is common to
define the two models by their action functionals:
\[
 E_{SG1} = \int \left\{
 \frac{|\nabla\psi|^2}{1-|\psi|^2} + (1-|\psi|^2) \right\} \text{d}^2x,
\]
and
\[
E_{SG2} = \int \left\{ \frac{|\nabla\psi|^2}{1-\tfrac{1}{2}|\psi|^2}
  + \tfrac{1}{2}(1-|\psi|^2)^2 \right\} \text{d}^2x,
  \]
respectively.

Recently there has been an upsurge of interest in the complex sine-Gordon
equations, motivated by the fact that
they define integrable perturbations of
 conformal field theories
\cite{Fateev,Bakas_Park,Brazhnikov,Bakas_Sonnenschein}. There is, however, yet
another reason for considering these systems more closely; out of all
vortex-bearing equations for one complex field, the complex sine-Gordon-1 and
-2 are the only equations whose vortex (and multivortex) solutions are
available in explicit analytic form \cite{coaxial,noncoaxial}. Consequently,
they provide a
 unique source
of insight into general properties of topological solitons on the plane. The
latter can be of value for a whole range of models like the Gross-Pitaevski and
easy-plane ferromagnet equations, where vortices are only available
numerically.

The (coaxial) multivortices of the complex sine-Gordon-2
have  been obtained via the
Schlesinger transformations of the fifth Painlev\'e equation \cite{coaxial}.
The procedure is cumbersome: even if the $(n-1)$-vortex is already
available,
the construction of the $n$-vortex solution requires applying the Schlesinger
transformation $2n$ times anew. On the contrary, there is an efficient
recursive procedure for the complex sine-Gordon-1, allowing a one-step
construction of its solution with vorticity $n$ provided the $(n-1)$-vortex
solution is known \cite{coaxial,noncoaxial}.
 The purpose of the present note
is to formulate a similar recursive procedure for the complex sine-Gordon-2.

 \textbf{2.} The coaxial $n$-vortex configuration
 has the form
 $\psi(r,\theta)= Q_n^{1/2}(r) e^{in\theta}$. Substituting
 this Ansatz into
\eqref{eq:csg2eqmotion} yields an equation for the
radial ``amplitude" $Q_n$ which we write as
\begin{multline}
\label{csg2Qgen}
\frac{d^2Q_n}{d{r}^2} + \frac{1}{r}\frac{dQ_n}{dr} + \frac{1-Q_n}{Q_n(Q_n -
2)}\left(\frac{dQ_n}{dr}\right)^2 + Q_n(1 - Q_n)(2 - Q_n)\\ + \frac{(a^2 -
b^2)Q_n}{r^2(2-Q_n)} + \frac{4a^2(1 - Q_n)}{r^2Q_n(2-Q_n)} + \frac{\gamma
Q_n(2-Q_n)}{2r}= 0,
\end{multline}
with $a = \gamma = 0$ and $b = -2n$.
 The last two terms
in (\ref{csg2Qgen}) being equal to
zero, this form may appear to be somewhat artificial.
However, there is an
 advantage in considering eq.(\ref{csg2Qgen}) with general
  $a$, $b$, and $\gamma$; namely,
the availability of transformations connecting
solutions with different sets of parameters.
Indeed, the change of variables \cite{coaxial}
\begin{equation}\label{eq:CSG2toPV}
  Q_n=\frac{2}{1-W}
\end{equation}
brings eq.\eqref{csg2Qgen} to
 the fifth Painlev\'{e} equation,
\begin{equation}\label{painV}
\frac{d^2W}{d{r}^{2}} + \frac{1}{r}\frac{dW}{dr} - \frac{3W -
1}{2W(W-1)}\left(\frac{dW}{dr}\right)^2 = \frac{(W-1)^2}{r^2}\left(\alpha W +
\frac{\beta}{W}\right) + \frac{\gamma W}{r} + \delta\frac{W(W+1)}{W-1},
\end{equation}
with $\alpha = \tfrac{1}{2}a^2$, $\beta = -\tfrac{1}{2}b^2$ and $\delta=2$.
 The Painlev\'{e}-V is covariant under the
Schlesinger transformation
\cite{Ablowitz,Gromak} which takes a solution $W$   with the
parameter values $a, b, \gamma$ and $\delta$ to a solution $\hat{W}$ of the
same equation with parameter values
\[
\hat{a} = \tfrac{1}{2}(a+b-1-\gamma/c),\quad \hat{b} =
\tfrac{1}{2}(a+b-1+\gamma/c),\quad \hat{\gamma} = c(b-a),
\]
and $\hat{\delta} = \delta$. Here $c$ is
one of the two values with   $c^2=-2\delta$; in our case we can set,
without loss of generality, $c=2i$.
Written in terms of $Q$ and $\hat{Q}=2(1-\hat{W})^{-1}$,
the direct and inverse
Schlesinger transformations have the form
\begin{subequations}
\label{schlesQ}
\begin{eqnarray}
\hat{Q} &=& 1
-\frac{i}{Q(Q-2)}
 \left[ \frac{dQ}{dr}+\frac{Q(a+b) - 2a}{r}\right],
 \label{schlesQa}\\
Q &=& 1 +\frac{i}{\hat{Q}(\hat{Q}-2)} \left[ \frac{d\hat{Q}}{dr}
 - \frac{(\hat{Q}-1)(a+b-1)}{r} +
\frac{i\gamma}{2r}\right].
 \label{schlesQb}
\end{eqnarray}
\end{subequations}

If $Q^{(0)}=Q_n$
is a solution of   eq.\eqref{csg2Qgen} with parameters $a^{(0)} = \gamma^{(0)} = 0$, $b^{(0)}
= -2n$,
then
applying transformation \eqref{schlesQa} we obtain a solution $\hat{Q} =
Q^{(1)}$ with parameters
\begin{eqnarray*}
a^{(1)} &=& \tfrac{1}{2}(a^{(0)}+b^{(0)}-1-\gamma^{(0)}/c)  = -n -\tfrac{1}{2},\\
b^{(1)} &=& \tfrac{1}{2}(a^{(0)}+b^{(0)}-1+\gamma^{(0)}/c)  = -n -\tfrac{1}{2},\\
\gamma^{(1)} &=& 2i(b^{(0)}-a^{(0)}) = -4in.
\end{eqnarray*}
Note that since $a^{(1)}$ and $\gamma^{(1)}$ are not zero and $b^{(1)}$
is not a negative even integer, $Q^{(1)}$ does not represent
the amplitude of any multivortex.
Using \eqref{schlesQa} again, this time with $Q = Q^{(1)}$ and $\hat{Q} =
Q^{(2)}$,
 yields
a solution $Q^{(2)}$ with parameters
\begin{eqnarray*}
a^{(2)} &=& \tfrac{1}{2}(a^{(1)}+b^{(1)}-1-\gamma^{(1)}/c)  = -1,\\
b^{(2)} &=& \tfrac{1}{2}(a^{(1)}+b^{(1)}-1+\gamma^{(1)}/c)  = -2n-1,\\
\gamma^{(2)} &=& 2i(b^{(1)}-a^{(1)}) = 0.
\end{eqnarray*}
Thus, the effect of the product transformation
$Q^{(0)} \to Q^{(2)}$ is to reduce both $a$ and
$b$ by one. Although $\gamma$ is now zero,
 $Q^{(2)}$
is still not a multivortex (since $a^{(2)}$ is nonzero).
 The crucial observation now is
 that \eqref{csg2Qgen} depends only on
the square of $a$; hence $Q^{(2)}$ is a solution of eq.\eqref{csg2Qgen} not
only for $a=-1$, $b=-2n-1$, but also for ${\tilde a}=+1$,
${\tilde b}=-2n-1$. Repeating
the above two
transformations will decrease both ${\tilde a}$ and ${\tilde b}$ by one more,
yielding $a^{(4)} = 0$,
$b^{(4)} = -2n-2$. The corresponding solution $Q^{(4)}$
will therefore be $Q_{n+1}$, the
multivortex with vorticity $n+1$. Thus, the transformation
$Q^{(0)} \to Q^{(4)}$, comprising a product of four Schlesinger
maps, is nothing but a vorticity-raising transformation  $Q_n \to Q_{n+1}$.

Starting with the trivial ``vortex"
$Q_0=1$ and applying the
 transformation $Q_n \to Q_{n+1}$ recursively,
one can  construct
multivortices of any desired
vorticity. It follows from the
form of the transformation that all
 $Q_n$'s will be rational functions of $r$.
The one-, two- and three-vortex amplitudes are:
\[
  Q_1 = \frac{r^2}{r^2+4},
  \]
  \[
  Q_2 = \frac{r^4(r^2+24)^2}{r^8+64r^6+1152r^4+9216r^2+36864},
\]
and
\[
  Q_3 = \frac{r^6(r^6+144r^4+5760r^2+92160)^2}{D_3},
 \]
 where
 \begin{eqnarray*}
  D_3 = r^{18} +
  324r^{16}+41472r^{14}+2820096r^{12}+114130944r^{10}+2919628800r^8 \\
  +50960793600r^6+61152952300r^4+4892236185600r^2+19568944742400.
\end{eqnarray*}
These solutions coincide with those constructed previously in
\cite{coaxial}.

 \textbf{3.} As another application of
the new vorticity-raising transformation, we construct one more class of
vortex-like solutions. Unlike the solutions
discussed above, these solutions of
the complex sine-Gordon-2 decay to their
asymptotic value in an
{\it oscillatory} fashion. (The asymptotic value is now  $\sqrt{2}$ not 1.)

We start with the radially symmetric solution
\cite{coaxial}  of the complex sine-Gordon-1 with
vorticity $n=2$:
\[
  \Phi_2 = -\frac{I_0I_2 - I_1^2}{I_0^2-I_1^2}.
\]
Here, $I_m=I_m(r)$ is the  modified Bessel function
of order $m$. The change of variables
\[
  \Phi_2 = \frac{1+W}{1-W}
\]
transforms it to a solution
\begin{equation}
  W_2(r) = \frac{I_0I_1-r(I_0^2 - I_1^2)}{I_0I_1}
  \label{W2}
\end{equation}
of the Painlev\'{e}-V, eq.\eqref{painV} with parameter values
\[
  \alpha = \tfrac{1}{2}, \quad \beta = -\tfrac{1}{2},\quad \gamma = 0, \quad
  \delta = -2.
\]
(In  \eqref{W2}, we simplified, using the standard recurrence relation
$I_0-I_2=2I_1/r$.)
Next, making the replacement $r \to ir$ amounts to replacing
$\gamma \to i\gamma$, $\delta \to -\delta$ in \eqref{painV}; hence
\[
  \tilde{W}_2(r)=W_2(ir) = \frac{J_0J_1-r(J_0^2 + J_1^2)}{J_0J_1}
\]
is a solution to the Painlev\'{e}-V with
\begin{equation}
  \alpha = \tfrac{1}{2}, \quad \beta = -\tfrac{1}{2},\quad \gamma = 0, \quad
  \delta = 2.
  \label{parameters}
\end{equation}
Here, $J_m=J_m(r)$ is the ordinary Bessel function of order $m$.

\begin{figure}
\input{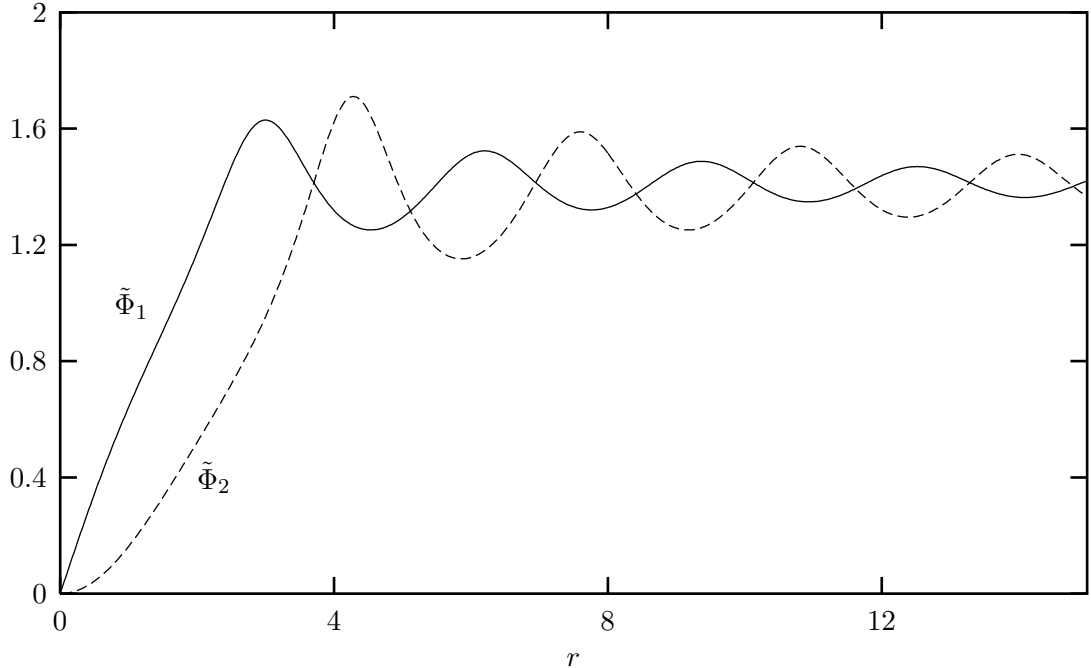}

  \caption{The moduli of the 1 and 2-vortex solutions \eqref{eq:csg2new}
  and \eqref{eq:csg2new2}. (As
    $r \rightarrow \infty$,
    $\tilde{\Phi}_1$ and $\tilde{\Phi}_2$ tend to $\sqrt{2}$.) }
    \label{fig:csg2newphi}
\end{figure}

For the parameter values in (\ref{parameters}), the change of
variables \eqref{eq:CSG2toPV} brings the Painlev\'e-V
to equation \eqref{csg2Qgen} with $a=1$ and $b=-1$.
 Thus we can use \eqref{eq:CSG2toPV} followed by
two applications of \eqref{schlesQa} to reduce both $a$ and $b$ by one (as
described in the previous section). The result is a solution to the complex
sine-Gordon-2 with $n=1$:
\begin{equation}\label{eq:csg2new}
  \tilde{Q}_1 = \frac{2\left[r(J_0^2+J_1^2) - J_0J_1\right]^2}{r^2(J_0^2+J_1^2)^2 +
  J_0^4}.
\end{equation}
Applying the vorticity-raising transformation to $\tilde{Q}_1$ gives the
formula for $\tilde{Q}_2$, the 2-vortex amplitude:
\begin{equation}\label{eq:csg2new2}
\tilde{Q}_2 =2\frac{P_2^2}{D_2},
\end{equation}
where
\begin{multline}
  P_2=
  4r^3J_0^4+8r^3J_0^2J_1^2+4r^3J_1^4-8r^2J_0^3J_1-8r^2J_0J_1^3+3rJ_0^4\\
    +2rJ_0^2J_1^2-5rJ_1^4-6J_0^3J_1+2J_0J_1^3,
    \nonumber
\end{multline}
and
\begin{multline}
  D_2 = 416r^4J_1^2J_0^6-128r^5J_0J_1^7+736r^4J_1^6J_0^2-384r^5J_0^3J_1^5+96r^6J_1^4J_0^4-12r^2J_1^2J_0^6\notag\\
   -96rJ_0^3J_1^5-320r^3J_0J_1^7-32rJ_0^5J_1^3+318r^2J_1^4J_0^4+64r^6J_1^2J_0^6-128r^5J_0^7J_1\notag\\
   -832r^3J_0^3J_1^5+260r^2J_1^6J_0^2-448r^3J_0^5J_1^3-384r^5J_0^5J_1^3+1008r^4J_1^4J_0^4 +64r^3J_0^7J_1\notag\\
   +r^2J_1^8+152r^4J_1^8+9r^2J_0^8+16r^6J_1^8+16r^6J_0^8+16J_1^4J_0^4-8r^4J_0^8+64r^6J_1^6J_0^2. \notag
\end{multline}
(The actual moduli ${\tilde{\Phi}}_1 = \sqrt{\tilde{Q}_1(r)}$
and ${\tilde{\Phi}}_2 = \sqrt{\tilde{Q}_2(r)}$
 are shown in figure \ref{fig:csg2newphi}.)
Proceeding recursively we can construct $\tilde{Q}_n$
with arbitrary $n$. For $n \ge 3$, the formulas become
intractable, so we only produce
the asymptotic behaviours, as $r \to \infty$:
\begin{equation}
\label{eq:asympt}
  \tilde{Q}_n \rightarrow 2 + (-1)^{n+1}\frac{2n}{r}\cos(2r) +
  \mathcal{O}\left(\frac{1}{r^2}\right).
\end{equation}
Eq.\eqref{eq:asympt} can be easily proved by induction.

This project was supported by the NRF of South Africa under
grant 2053723, by the Johnson Bequst Fund and the URC
of the University of Cape Town.

\end{document}